\documentclass[aps,prl,twocolumn,superscriptaddress,nofootinbib,showpacs]{revtex4}

\usepackage{slashed}
\usepackage{epsfig,latexsym,cancel,amssymb,amsmath,verbatim,mathrsfs}
\usepackage{color}
\usepackage{graphicx}

\usepackage{epsfig}
\usepackage{amsmath}
\usepackage{pstricks}
\usepackage{bbold}
\usepackage{slashed}
\usepackage{amssymb}
\usepackage{graphicx}
\usepackage{hyperref}
\usepackage{verbatim}
\usepackage{multirow}
\usepackage{chemarrow}

\def\ra{\rightarrow}

\def\wt{\widetilde}

\def\ld{\lambda}
\def\f{\frac}
\newcommand{\be}{\begin{equation}}
\newcommand{\ee}{\end{equation}}
\newcommand{\bea}{\begin{eqnarray}}
\newcommand{\eea}{\end{eqnarray}}
\newcommand{\ba}{\begin{array}}
\newcommand{\ea}{\end{array}}

\long\def\symbolfootnote[#1]#2{\begingroup%
\def\thefootnote{\fnsymbol{footnote}}\footnote[#1]{#2}\endgroup}

\newcommand{\beq}{\begin{equation}}
\newcommand{\eeq}{\end{equation}}

\newcommand{\ET}{\mbox{$\not \hspace{-0.10cm} E_T$ }}
\newcommand{\PT}{\mbox{$\not \hspace{-0.10cm} p_T$ }}

\begin{document}

\title{New Physics Opportunities in the Boosted Di- Higgs plus \ET Signature}

\author{Zhaofeng Kang}
\email[E-mail: ]{zhaofengkang@gmail.com}
\affiliation{School of Physics, Korea Institute for Advanced Study,
Seoul 130-722, Korea}

\author{P. Ko}
\email[E-mail: ]{pko@kias.re.kr}
\affiliation{School of Physics, Korea Institute for Advanced Study,
Seoul 130-722, Korea}

\author{Jinmian Li}
\email[E-mail: ]{phyljm@gmail.com}
\affiliation{ARC Centre of Excellence for Particle Physics at the Terascale, Department of Physics, University of Adelaide, Adelaide, SA 5005, Australia}

\date{\today}% It is always \today, today,

\begin{abstract}

The  Higgs field in the standard model (SM) may couple to new physics sectors related with dark matter 
and/or massive neutrinos. In this paper we propose a novel signature, the boosted di-Higgs boson plus \ET (which is either a dark matter or neutrino), to probe those new physics sectors. In a large class of models, in particular the supersymmetric SMs and low scale seesaw mechanisms, this signature can play a key role. The signature has clear background,  and at the $\sqrt s=$14 TeV high luminosity (HL-)LHC, we can probe it with production rate as low as $\sim$ 0.1 fb.  We apply it to benchmark models, supersymmetry in the bino-Higgsino limit, the canonical seesaw model and little  Higgs model, finding that masses of  Higgsino, right-handed neutrino and heavy vector boson can be probed up to $\sim$ 500 GeV, 650 GeV and 900 GeV, respectively.

\end{abstract}

\maketitle

\noindent {\bf{New physics below the iceberg}}  Discovery of Higgs boson at the large hadronic collider (LHC)~\cite{Higgs:dis} completes the standard model (SM). At the same time, however, it opens a new era for particle physics: this new resonance might be just a small tip of a big iceberg, and below it could hide a mystery of new world. Looking around this small tip, we may find clues for new physics. Actually, we do have convincing arguments to support this belief from several motivations for new physics beyond the SM (BSM). 

The most common theoretical argument for new physics is from the notorious gauge hierarchy problem caused by the quadratic divergence of (Higgs mass)$^2$ parameter. Solutions to this problem introduce new particles coupled to the SM Higgs field. For instance, in the supersymmetric SMs (SSMs) the Higgs doublet $H_u$ (along with $H_d$ by virtue of anomaly cancelation) participate in quite a few interactions. The second argument for new physics is from dark matter (DM), whose interactions with the visible sector may be through the SM Higgs boson $h$ (Higgs portal DM). Last but not the least, neutrinos in the renormalizable SM are massless in conflict with the observation. Mechanisms, such as various seesaw mechanisms, to generate neutrino masses may again introduce new couplings to $h$. It is well expected that these new interactions of $h$ could contribute to the production of $h$ (plus something else) at LHC. In this letter we will concentrate on the di-Higgs production, which is common in these mentioned BSM contexts, has a bright prospect at LHC and moreover provides a new angle on di-Higgs physics.

\noindent {\bf{BSM with boosted di-Higgs bosons plus \ET}}  Di-Higgs boson search is one of the focuses in the upcoming LHC run, aiming at examining the  Higgs potential~\cite{Gupta:2013zza,di Higgs}. On the other hand, on top of the di-Higgs, a large class of BSM models can produce associated objects $X$, e.g., $X$ is a missing particle at LHC such as DM or neutrino. Moreover, both Higgs bosons may be boosted if they are produced from heavy particle decay. These additional features could greatly enhance the di-Higgs searches at a hadronic collider. In the following we select two well-known BSM models where $X$ is a DM and neutrino, respectively.

The first example is the neutralino sector of the minimal SSM (MSSM) with bino-Higgsino  accessible at LHC only, $\psi=(\wt B,\wt H_d^0,\wt H_u^0)$.~\footnote{The gravitino-Higgsino system with very light (says $\sim$keV) gravitino from the low scale supersymmetry-breaking models fit this scenario better. We leave it for a future study.} Since we are interested in boosted di-Higgs, we assume such a mass hierarchy: Higgsinos $\wt H_{u,d}^0$ (with mass $\mu \sim 500$ GeV) are much heavier than bino $\wt B$ (with mass $\lesssim 100$ GeV). The mass eigenstates $\chi_{1,2,3}$, respectively having masses $M_{1,2,3}$ in ascending order, are related to $\psi$ via $\psi_i=Z_\psi^{ij}\chi_j$. $\chi_2$ and $\chi_3$ are Higgsino-like, constituting a pseudo Dirac fermion pair with masse splitting of a few GeVs; $\chi_1$ is treated as massless for the moment. Electroweak gauge interactions of these particles are described by 
\begin{align}\label{}
{\cal L}_\chi=&\f{{g_2}}{2}\tan\theta_w\bar\chi_i a^h_{ij}\chi_j h+\f{g_2}{4\cos\theta_w}\bar\chi_i  a^z_{ij}\gamma_5\gamma_\mu \chi_j Z^\mu,
\end{align}
with $\theta_w$ the Weinberg angle and $a_{ij}^h=Z_\psi^{2j}Z_\psi^{1i}$, $a_{ij}^z=Z_\psi^{3i}Z_\psi^{3j}-Z_\psi^{2i}Z_\psi^{2j}$. We have assumed CP-invariance and an exact decoupling 
between two Higgs doublets. Note that the diagonal couplings $a_{ii}^z$ are suppressed by the small mass splitting between $M_2$ and $M_3$. 

The interesting signature is produced along the chain $ p p\ra Z^*\ra\chi_2 (\ra \chi_1 h)\chi_3 (\ra\chi_1 h)$. In the high energy limit $M^2_{2,3}-M_1^2\gg m_Z^2$,  we can apply the Goldstone equivalence theorem and obtain the following equations for decay branching ratios of $\chi_{2,3}$ (We refer to~\cite{Jung:2014bda} for relevant discussions on this point):
\begin{align}\label{}
&\f{\Gamma(\chi_2\ra h+\chi_1)}{\Gamma(\chi_3\ra Z+\chi_1)}\approx \f{\Gamma(\chi_2\ra Z+\chi_1)}{\Gamma(\chi_3\ra h+\chi_1)}.
\end{align}
This theorem will be used frequently in this paper. In order to approach the maximal branching ratio of the di-Higgs boson channel, $\Gamma(\chi_i\ra h+\chi_1)\approx\Gamma(\chi_i\ra Z+\chi_1)=50\%$ is favored. And it 
holds when the Higgs sector gives a large $\tan\beta$ limit and moreover $|\mu|\approx |M_{2,3}|\gg M_1$. In this limit the cross section of di-Higgs plus \ET can reach 1.2 fb for 500 GeV Higgsinos.

The second example is the heavy right-handed neutrino (RHN) $N$ in the type-I seesaw mechanism implemented in the local $B-L$ model~\cite{Basso:2008iv}. The minimal model is described by the following Lagrangian:
\begin{align}\label{seesaw}
-{\cal L}_{N}=&g_{B-L}\bar N\gamma_\mu  P_R NZ_{B-L}^\mu+y_{N}\bar \ell \wt H P_R N\cr
&+\f{1}{2}\ld_N\Phi\bar N^c N+h.c.+V(\Phi,H),
\end{align}  
where we consider only one family of RHN for simplicity. $N$ carries one unit of $B-L$ charge and gains Majorana mass from the coupling to the $B-L$ Higgs field $\Phi$, which develops VEV $v_\phi$ from the potential $V(\Phi,H)$ and breaks $U(1)_{B-L}$ at the TeV scale. The term $|\Phi|^2|H|^2$ in $V(\Phi,H)$ mixes $\Phi$ and $H^0$, the neutral component of SM doublet $H$, leading to two Higgs bosons in the mass eigensate:
\begin{align}\label{}
h=\cos\theta \,H^0_R+\sin\theta \Phi_R,\,\,\phi=-\sin\theta \,H^0_R+\cos\theta \Phi_R.
\end{align}
We have decomposed $\Phi=(v_\phi+\Phi_R+i\Phi_I)/\sqrt{2}$ similarly for $H^0$. The current data tells that $h$ is quite SM-like, and the measured Higgs signal strength imposes an upper bound on the mixng angle $\sin\theta\lesssim0.34$ at the 95\% CL~\cite{Lopez-Val:2014jva}. For such a small mixing angle, the SM-like Higgs boson mass squared can be approximated as
\begin{align}\label{}
m_h^2\approx \ld_h v^2-{\sin^2\theta}m_\phi^2,
\end{align}
where $\ld_h$ is the usual  Higgs  quartic self coupling.  Let us mention that the heavier $\phi$ and the larger $\theta$ are good for solving the metastability problem of  Higgs potential near the Plank scale, which is attributed to the relatively small $\ld_h= \ld_{\rm SM}\approx0.26$. For example, for $m_\phi=1.0$ TeV and $\sin \theta=0.3$, we now need $\ld_h\approx 1.7$. Note that such discussions can be easily generalized to other models. 

The heavy $N$ can be abundantly pair-produced via the resonance $Z_{B-L}/\phi$: $pp\ra Z_{B-L}/\phi\ra NN$.  The $Z_{B-L}$ channel is suppressed by the small branching ratio Br$(Z_{B-L}\ra NN)\approx 3\%$. In addition to that, it is strongly constrained by the di-lepton resonance search $pp\ra Z_{B-L}\ra \bar \ell \ell$ at LHC~\cite{Aad:2014cka}. The $\phi$ channel is due to the gluon-gluon fusion production of $\phi$, described by the usual dimension-five operator for $\phi\rightarrow gg$:
\begin{align}\label{Effective}
\f{1}{4}C_{gg\phi} G_{\mu\nu}^aG^{a\,\mu\nu}\f{\phi}{v},
\end{align}
with $C_{gg\phi}= -\sin\theta \,\alpha_s/2\pi $. However, this effective description becomes invalid when $m_\phi$ is much heavier than the top quark mass. Therefore, in the actual LHC analysis, we will utilize the results from the CERN Yellow Report~\cite{higgs14} which takes into account the finite top quark mass effect. For $\ld_N\sim1$ one can naturally expect Br$(\phi\ra NN)\sim 100\%$. The RHN pair is followed by the decay $N\ra \nu_L h$ with $\nu_L$ the active neutrino, giving rise to the boosted di-Higgs plus \ET signature. Because RHN is heavy, the branching ratios satisfy the relations~\cite{Huitu:2008gf,Basso:2008iv}:
\begin{align}\label{}
{\Gamma (N\ra \nu_L h)}\approx \Gamma (N\ra \nu_L Z)=\f{1}{2}\Gamma (N\ra \ell W),
\end{align}
which again follows from the equivalence theorem. For $M_N=0.5$ TeV, one can reach a signal cross section 1.5 fb given that $m_\phi$ is about 1 TeV.

Comments are in order. The conventional single RHN (below the weak scale) production counts on sizable mixings between the sterile and active neutrinos~\cite{RHN:mix,Deppisch:2015qwa}, which would be extremely small in generic cases. But probably RHN has gauge and/or Yukawa couplings, e.g., in the local $B-L$ models, and thus the pair production of heavy RHN is hopeful at 14 TeV LHC or other future colliders~\cite{RHN:100}. In particular, to our knowledge, using a scalar resonance is novel in RHN production and may offer the unique chance to probe type-I seesaw mechanism.

\noindent {\bf{Simplified models for $2h+$\ET}}  There are many other well motivated BSM models that can produce this signature; see an incomplete collection in the supplement material. Therefore, before we head toward the detailed collider study, this would be the right place to develop simplified models which could be used for a more general study of this signature.  

Two classes of models are of interest. Let us start with the first class, where the missing particle is DM, or more widely a  neutral particle stable at the collider time scale. The dark sector is supposed to consist of several dark states; we will use this term to genetically refer to particles undiscovered. And we will consider two dark states for concreteness. Then their interactions with the  Higgs boson would take the following forms:
\begin{align}\label{effective}
\ld_f h\chi_1\chi_2,\quad \mu_s hS_1S_2,\quad g_Z h Z_2^\mu (Z_1)_\mu.
\end{align}
($\chi_1$, $S_1$, $Z_1$) is a (Majorana fermion, real scalar, vector boson) DM with negligible mass; the heavy dark state $\chi_2$, etc., is in the sub-TeV mass region. Similarly, the second class with neutrino as the missing particle can be built. RHN-like heavy state $N$ couple to neutrino and  Higgs boson via the effective operator
\begin{align}\label{}
\ld_\nu h N \nu_L.
\end{align}
With these, the event topology boosted $2h+$\ET is generated through pair production of heavy dark states which decay into DM/neutrino plus  Higgs bosons. 

%~\footnote{Such a father particle will be collectively denoted as $F$ hereafter.}

In the simplified models we do not specify the production mechanism for the father particle $F$ such as $\chi_2$ since it is fairly model dependent. We can consider three possibilities for $F$ productions. Firstly, the father particle $F$ itself such as the Higgsino has electroweak interactions so that the $F$ pair can be produced via Drell-Yan processes. This mechanism does not involve extra particles and couplings. Secondly, in the presence of a new particle $Y$ coupling to light quarks $q$ as $\sim q Y F$ (just schematically), $F$ can be pair-produced via exchanging $Y$ in the $t-$channel. In the supplement, the vector $F$ pair production in the little  Higgs model is discussed in this way. Last but not the least, if $Y$ has interactions like $\sim Y q  q/ GG +Y FF $, then the $F$ pair can be resonantly produced. But as mentioned before, this contribution will be stringently restricted if $Y$ also couples significantly to leptons. The latter two possibilities may 
provide more effective production mechanisms than the first one, provided that the mass and 
couplings of $Y$ (to quarks and $F$) are proper.  In the following studies, a proper production mechanism for the $F$ pair will be assumed.

\noindent {\bf{Boosted 2$h+\ET$ at the 14 TeV LHC}}  The hadronic $4b$ mode is dominant in the di-Higgs decay, and becomes relieved from the huge QCD backgrounds (BGs) in the boosted region. For example, D. Lima {\it et al.} show that with the help of the boosted di-Higgs channel one can finally reach $1.2 \times \lambda_{\rm SM}$ at 14 TeV HL-LHC~\cite{deLima:2014dta}.~\footnote{The ATLAS search on data set of 8 TeV 19.5 fb$^{-1}$ shows that the current search sensitivity to the boosted di-Higgs channel is around an order of magnitude above the SM di-Higgs production rate~\cite{Aad:2015uka}, without using the substructure technique.} 
This channel was believed to be not promising~\cite{2h1,2h2} outside the boosted Higgs region. In some new physics scenarios, where the di-Higgs is produced from a heavy resonance decay, this channel is even more remarkable~\cite{Cooper:2013kia,Gouzevitch:2013qca,Khachatryan:2015yea}. Other related studies involving boosted (maybe extra) Higgs boson(s) are also motivated in various contexts~\cite{Kang:2013rj,Cooper:2013kia,b2h,Chen:2014dma,Chakraborty:2015xia}. In our signature, the (highly) boosted di-Higgs is further strengthened by a large missing energy, that could lead to the much earlier discovery of 2$h$$+\ET$ than the previous signatures without \ET.~\footnote{CMS searched for this signature without using boosted Higgs-tagging\cite{CMS-PAS-SUS-13-022}. They required that all $4b$'s have to be resolved.}

The BGs of 2$h$$+\ET$ are similar to those of  the di-Higgs signature: the irreducible BGs (dominated by QCD $4b$ and $Zb \bar{b}$) and the reducible BGs (dominated by the semi-leptonic $t\bar{t}$ pair); $4j$ with $j=g,u,d,s,c$ can be mis-tagged as $4b$, but the cross section is $\lesssim10^{-3}\times \sigma(4b)$ and thus negligible. In the irreducible BGs, \ET is due to the limited detector resolution, and the distribution of \ET depends on the detector setup: e.g., at 7 TeV LHC it respects a Gaussian distribution with the central value $\sim 0.5\sqrt{\sum p_T}$ GeV~\cite{Aad:2012re}. While in the semi-leptonic $t\bar{t}$ BG, light flavor jets can be mis-tagged as $b$-jets and at the same time the leptons may be missed, which happens, besides owing to the limited lepton tagging efficiencies (especially for $\tau$), if the leptons go outside the kinematic region, i.e., $p_T(l)>10$ GeV and $|\eta|<2.5$, or they are not isolated, i.e., the scalar $p_T$ sum of particles in the vicinity ($\Delta R<0.5$) of the lepton is greater than $10\%\times p_T(l)$. MadGraph5\_aMC@NLO~\cite{Alwall:2011uj} is used to calculate the LO cross sections for 4$b$ (861 pb) and $Zb\bar{b}$ (109 pb at NLO after multiplying a constant $K$-factor 1.4~\cite{Campbell:2003hd}); the bottom quarks in these BGs are required to have $p_T(b) > 20$ GeV, $|\eta(b)|<2.5$ and $\Delta R(b,b)>0.4$. The NNLO cross section of the semi-leptonic $t\bar{t}$ including all lepton flavors can be found in Ref.~\cite{Ahrens:2011px}, 382 pb; no cuts are imposed on the top decay products.

We take heavy RHN pair production via the $\phi-$resonance as the representative signal process, setting $m_{\phi} = 2 M_{N}+30$ GeV and Br$(N\to h \nu_L)=1$. FeynRules~\cite{Alloul:2013bka} is used for generating the UFO model files of the model~\cite{Basso:2011na}. The signal (and BGs) events are generated by MadGraph5\_aMC@NLO, and passed to Pythia6~\cite{Sjostrand:2006za} for SM particles decay, parton showering and hadronization. The detector effects are simulated with Delphes3~\cite{deFavereau:2013fsa}, where we choose the default ATLAS detector setup. The $b$-tagging efficiency is set to 0.7 with mis-$b$-tagging rates for $c$- and light-flavored jets assumed to be 0.1 and 0.015~\cite{CMS-PAS-BTV-13-001}, respectively. Then, the particle flow information from Delphes3 is analyzed using fastjet~\cite{Cacciari:2011ma}.

Comments on the generalization of the above representative model are in orders. Firstly, the invisible particle 
is assumed to be massless. However, our results can be generalized to the massive invisible particle case (\ET refers to the measurable \PT in this case), because the variables in our analysis actually are merely sensitive to the mass difference between the invisible particle and the father particle. Secondly, we have chosen the Yukawa coupling of $\phi$ such that the narrow width approximation holds ($\Gamma_\phi\lesssim 0.1m_\phi$); moreover, $m_{\phi}-2 M_{N}$ is small. So, it is justified to conclude that the kinematic properties of the RHN pair produced here are similar to those of the $F$ pair which is produced in other ways through an off-shell $s$-channel mediator or a $t$-channel mediator.

In selecting the signal events, lepton veto is imposed first. Next, the Cambridge/Aachen algorithm~\cite{CA} with an appropriate cone size $R_{\rm fat}$ is used to recluster fat-jets. Then the BDRS algorithm~\cite{BDRS} is applied to resolve their substructure.  More concretely, a Higgs jet candidate during the declustering should have large mass drop $\mu = \frac{m_{j_1}}{m_j} < 0.67$, and not too asymmetric mass splitting $y = \frac{\min (p^2_{T, j_1} , p^2_{T,j_2})}{m^2_j} \Delta R^2_{j_1, j_2} > 0.09$. After filtering, the three hardest subjets inside the fat-jet, 
which are reconstructed by the anti-kt algorithm with $R_{filt}=\min(0.3,R_{b\bar{b}}/2)$, are identified as the ingredients of the Higgs jet candidate; also the two leading subjets are required to be $b$-tagged.

The separation angle between $b$ and $\bar b$ from Higgs decay, namely the Higgs jet cone size, depends on 
the energy of Higgs boson. For each RHN mass $M_N$, we scan $R_{\rm fat} \in [0.5, 3]$ with step size 0.1 and 
find $R_{\rm fat}^{max}$, that retains most of the signal events whose leading two fat-jets pass the BDRS 
Higgs-tagging criteria and the cut $m_{\rm  Higgs} \in [85,160]$ GeV (A more refined mass interval for a given 
$M_N$ is given in Table II.). In Table~\ref{bench} we show the options of $R_{\rm fat}^{max}$ for different values 
of $M_N$. The fraction of retained signal events (selection efficiency) and cross sections of BGs after 
Higgs-tagging are also displayed. As expected, a smaller $R_{\rm fat}^{max}$ should be adopted for a heavier 
$N$. It then greatly reduces the probability that two close QCD jets mimic the Higgs jet, thus suppressing BGs (especially the QCD BG from $4b$'s).  
 \begin{table}[htb]
 \begin{center}
  \begin{tabular}{|c|c|c|c|c|c|} \hline
 $M_N$ (GeV)  &$R_{\text{fat}}^{\text{max}}$  & {$S$} & $t\bar{t}$ (fb) & $Zb\bar{b}$(fb) & QCD(fb) \\ \hline
  200 &  2.0    & 2.1\%  & 5.2 & 13.9 & 591.8 \\
  300 &  1.8  &  2.5\% & 3.9 & 9.2 & 399.4 \\
  400 & 1.6 & 3.2\% & 2.4 &5.6 & 241.7 \\
  500 &  1.6  &  3.9\% & 2.4  & 5.6 & 241.7 \\
  700 &  1.4  & 5.4\% & 1.3 & 2.9 & 117.1 \\
  1000 &  1.0  & 6.8\% &  0.46 & 0.41 & 14.5 \\
  2000 &  0.8  & 8.8\% & 0.19 & 0.1 & 3.2 \\ \hline
  \end{tabular}
  \end{center}
\caption{\label{bench} The selection efficiency $S$ and cross sections of BGs after applying Higgs-tagging which takes $m_{\rm  Higgs} \in [85,160]$ GeV and $R_{\rm fat} = R_{\text{fat}}^{\text{max}}$.}
\end{table}

There are some other powerful discriminators in our analysis, such as $\ET$ and the reconstructed two Higgs 
boson masses. Requiring $\ET \gtrsim 100$ GeV after Higgs-tagging leaves the semi-leptonic $t\bar{t}$ as the dominant BG.   In addition, since we consider pair production of heavy particles, the $M_{T_2}$ variable~\cite{MT2}, which reflects mass difference between the father particle and its invisible daughter, is quite useful. According to these variables, we design four signal regions (SR) which are suitable for different $M_N$; see Table~\ref{cuts} 
where the name of SR is indicated by the RHN mass. The cuts of each SR are optimized with respect to the corresponding benchmark point. Note that SR500 will be used for $M_N$ above 500 GeV because its BGs 
already become negligible. 
\begin{table}[htb]
 \begin{center}
  \begin{tabular}{|c|c|c|c|c|} \hline
signal region &  SR200 & SR300 & SR400 & SR500  \\ \hline
 \multirow{2}{*}{Selection cuts} &  \multicolumn{4}{|c|}{Lepton veto \& tau veto} \\ 
  &  \multicolumn{4}{|c|}{Two  Higgs tagged jets} \\ \hline
  $\slashed{E}_T$/GeV & $>100$ & $>100$ & $>200$ & $>300$  \\ \hline
  $m_{h_1}$/GeV & [90,150] & [90,150] & [90,150] & [90,150] \\ \hline
  $m_{h_2}$/GeV & [80,140] & [70,150] &  [80,140] & [80,140] \\ \hline
  $M_{T_2}(h_1,h_2)$ & $>130$ & $>170$ & $>215$ & $>300$ \\ \hline
  \end{tabular}
   \end{center}
  \caption{\label{cuts} Optimized cuts for the four signal regions.}
\end{table}
 
\begin{figure}[htb]
 \begin{center}
     \includegraphics[width=0.4\textwidth]{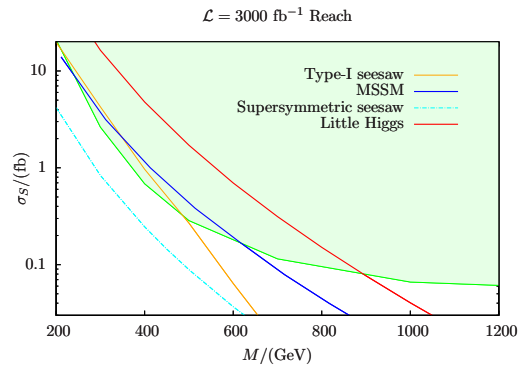}
  \end{center}
    \caption{\label{exclu}  Based on the representative signal process, the shaded region can be probed with 
3-$\sigma$ significance at 14 TeV 3000 $fb^{-1}$ LHC. We display the cross sections of di-Higgs$+\ET$ in four models: (I) seesaw (orange), (II) MSSM (blue), (III) supersymmetric seesaw (cyan) and (IV)  little  Higgs (red) 
models.} 
\end{figure}

\noindent {\bf{How far can we reach?}}  For a given RHN mass, the signal cross section that provides at least 3-$\sigma$ significance, i.e., $S/ \sqrt{S+B}\geq 3$, is
\begin{align}
\sigma_S \geq \frac{1}{\epsilon_S \mathcal{L}} \frac{9+\sqrt{81+36\times \sigma_B \epsilon_B 
\mathcal{L}}}{2}~,~
\end{align}
where $\mathcal{L}$ is the integrated luminosity and $\epsilon_{S,B}$ are cuts efficiencies for the signal and 
BGs, respectively. In Fig.~\ref{exclu} we show the reach limit of our representative signal process in the light 
green shaded region.  One can see that the production rate as low as ${\cal O}(0.1)$ fb can be reached for 
$M_N >500$ GeV.

Now we explain the above model-independent reach in several benchmark models, where the kinematic properties of the father particle ($F$) pair are similar to those of the RHN pair in the representative signal process. They are: (I) the neutrino system in the Type-I seesaw that presents a scalar resonance $\phi$ with a large mass $m_{\phi} \gtrsim 2 M_{N}$ (For such heavy $\phi$, its production rate is obtained from Ref.~\cite{higgs14} rather than Eq.~(\ref{Effective}).), the mixing angle~\footnote{Converting the signal reach to the limit on the mixing angle for $\phi$, we get $\sin\theta\gtrsim 0.23$ for $m_\phi$ around 800 GeV. But it is can be improved much at the 100 TeV collider~\cite{Kang:2015uoc}.} $\sin\theta=0.3$ and Br($\phi \to NN)=1$; (II) the bino-Higgsino system in MSSM with Br$(\tilde{H_{u,d}^0} \to \tilde{B} h)= 50\%$; (III) the sneutrino system $\wt\nu_L- \wt N$ in the supersymmetric seesaw~\cite{Dirac:snu} with the left-handed sneutrino $\wt\nu_L$ produced via the Drell-Yan process and Br$(\wt\nu_L\ra \wt N+h)=50\%$; (IV) the massive vector boson system $Z_H-A_H$ in the little Higgs model~\cite{Arkani,Hubisz:2004ft} with $Z_H$ produced by the $t$-channel exchange of a quark partner and Br$(Z_H\ra A_H+h)=100\%$. We put the details of the later two models in the supplement. The  3-$\sigma$ discovery lines are labelled in Fig.~\ref{exclu}. We observe that (RHN,  Higgsinos, $Z_H$) with mass up to $\sim (0.50, 0.65, 0.9)$ TeV can be probed at 3-$\sigma$ level, whereas the sneutrino case can be hardly probed. Note that the invisible particle is assumed to be massless, and the violation of this assumption will lead to a shift of these curves toward left, roughly by the invisible particle mass.

\noindent {\bf{Conclusion}}  In the light of a large class of BSM models, we propose a noble signature, namely the boosted di-Higgs plus \ET to probe new physics. Good prospects of models like MSSM, type-I seesaw and little Higgs are demonstrated,  finding that masses of  Higgsino, right-handed neutrino and heavy vector boson can be probed up to $\sim$ 500 GeV, 650 GeV and 900 GeV, respectively.

\noindent {\bf{Acknowledgements}} 
%We would like to thank  for helpful discussions. 
This work is supported in part by National Research Foundation of Korea (NRF) Research 
Grant NRF-2015R1A2A1A05001869, and by the NRF grant funded by the Korea 
government (MSIP) (No. 2009-0083526) through  Korea Neutrino Research Center 
at Seoul National University (PK).

\end{document}